\documentclass[fleqn,usenatbib]{mnras}

\usepackage{newtxtext,newtxmath}
\usepackage[T1]{fontenc}

\DeclareRobustCommand{\VAN}[3]{#2}
\let\VANthebibliography\thebibliography
\def\thebibliography{\DeclareRobustCommand{\VAN}[3]{##3}\VANthebibliography}

\usepackage{graphicx}	% Including figure files
\usepackage{amsmath}	% Advanced maths commands

\title[JWST Reveals an Association of RSGs with OB Stars]{JWST Reveals a Galaxy-Wide Association of Red Supergiants with OB Stars in NGC~5584}

\author[Min Dai et al.]{
Min Dai$^{1,2,3}$,\thanks{E-mail: shuwang@nao.cas.cn, bjiang@bnu.edu.cn}
Shu Wang$^{4}$
and Biwei Jiang$^{2,3}$
\\
$^{1}$School of Physics and Astronomy, China West Normal University, No. 1 Shida Road, Nanchong 637009, People's Republic of China\\
$^{2}$School of Physics and Astronomy, Beijing Normal University, Beijing 100875, People's Republic of China\\
$^{3}$Institute for Frontiers in Astronomy and Astrophysics, Beijing Normal University, Beijing 102206, People's Republic of China\\
$^{4}$CAS Key Laboratory of Optical Astronomy, National Astronomical Observatories, Chinese Academy of Sciences, Beijing 100101, People's Republic of China
}

\date{Accepted XXX. Received YYY; in original form ZZZ}

\pubyear{2026}

\begin{document}
\label{firstpage}
\pagerange{\pageref{firstpage}--\pageref{lastpage}}
\maketitle

% Abstract of the paper
\begin{abstract}
The spatial relationship between young and evolved massive-star tracers offers a means of probing recent star formation across galactic disks. We combine JWST/NIRCam, $Swift$/UVOT UVW2, and GALEX FUV/NUV images to investigate the spatial association between red-supergiant (RSG) candidates and UV-selected star-forming complexes (SFCs) in the spiral galaxy NGC~5584. We identify 5,310 RSG candidates using colour--magnitude criteria and define 106 UV-selected SFCs from the UVOT/UVW2 morphology, with their FUV and NUV luminosities measured independently using the GALEX images. We find that 82\% of the UV-selected SFCs overlap significant RSG overdensities, indicating a galaxy-wide statistical association between UV-bright SFCs and the evolved massive-star population. Among the 93 SFCs with reliable JWST coverage, the UV-$\beta$-corrected FUV and NUV luminosities correlate tightly with the JWST/NIRCam luminosities, with Pearson correlation coefficients of $r=0.962$--$0.971$. MIST isochrone comparisons for 43 RSG-rich SFCs indicate a characteristic age of $\log(\mathrm{age/yr})\simeq 7.16$. These results show that RSG overdensities and UV-selected SFCs trace related but distinct phases of recent massive-star formation. Their spatial correspondence is statistical rather than one-to-one and likely reflects differences in stellar age, dust extinction, spatial resolution, aperture coverage, and local star-formation history.
\end{abstract}

% Select between one and six entries from the list of approved keywords.
% Don't make up new ones.

%%%%%%%%%%%%%%%%%%%%%%%%%%%%%%%%%%%%%%%%%%%%%%%%%%

%%%%%%%%%%%%%%%%% BODY OF PAPER %%%%%%%%%%%%%%%%%%
\section{Introduction}\label{intro}
Massive stars (with initial masses $>\sim8\nobreak\,\mathrm{M}_{\odot}$) are the primary drivers of cosmic evolution \citep{2005ARA&A..43..769V, 2012ARA&A..50..531K, 2019A&ARv..27....3M}. Through their powerful stellar winds, binary interactions, and core-collapse supernova (CCSN) explosions, they regulate the energy and chemical budgets of their host galaxies and shape the surrounding interstellar medium. Understanding their life cycles is therefore crucial for interpreting the formation and evolution of galaxies over cosmological timescales. 

A key aspect of massive-star evolution is the relation between hot, short-lived massive stars and the red supergiant (RSG) phase. Massive stars spend approximately 1-20 Myr on the main sequence before the most massive members evolve through later phases that may include an RSG stage \citep{2012A&A...537A.146E,2016ApJ...823..102C}.
RSG candidates and ultraviolet-bright star-forming complexes (SFCs; hereafter also referred to as clumps) therefore sample overlapping but non-identical intervals of recent star formation. Their spatial distributions can be compared to characterize how young and somewhat more evolved massive-star tracers are arranged across a galactic disk. Spatial correspondence alone, however, neither demonstrates triggered star formation nor establishes a sequence of distinct stellar generations.

Galaxy-wide comparisons of resolved RSG candidates with ultraviolet-selected SFCs remain rare. Studies within the Milky Way have provided foundational insights into the spatial organization of young massive-star populations and OB associations \citep{1947esa..book.....A, 2003ARA&A..41...15M, 2020A&A...644A..62C}, while observations of external galaxies have reported qualitative associations between RSGs and spiral-arm structures \citep{2015A&A...578A...3G, 2021ApJ...922..177M}. However, stellar crowding and the large difference in spatial resolution between ultraviolet and near-infrared data complicate a consistent galaxy-wide analysis. Quantifying the statistical correspondence between these tracers provides an empirical description of recent massive-star populations across a full galactic disk, while departures from one-to-one correspondence constrain the importance of age, dust, crowding, and spatial scale.

The nearby spiral galaxy NGC~5584 provides a valuable laboratory for investigating the connection between young and evolved massive-star populations on galaxy-wide scales. NGC~5584 is a late-type, star-forming spiral galaxy with a weak bar and a patchy, flocculent disk morphology, with recent star formation distributed across multiple spiral-arm segments and SFCs. The distance to NGC 5584 is approximately
23.3\nobreak\,Mpc \citep{2024ApJ...962L..17R},
corresponding to a physical scale of
113\nobreak\,pc\nobreak\,arcsec$^{-1}$. Basic properties of NGC~5584 are summarized in Table~\ref{tab:ngc5584_properties}. Its relatively low inclination provides a favorable view of the spatial distribution of stellar populations and SFCs across a range of galactic environments. At this distance, however, stellar crowding remains a significant observational challenge, especially in regions of active star formation, making it difficult to construct complete and reliable samples of individual massive stars using optical data alone.

The sensitivity and angular resolution of \textit{JWST}/NIRCam enable the identification of luminous evolved massive-star candidates, including RSG candidates, across the crowded disk of NGC~5584. To relate these evolved stellar populations to recent star formation, we combine the near-infrared \textit{JWST} observations with ultraviolet imaging from the \textit{Swift} Ultra-Violet/Optical Telescope (UVOT) and GALEX. The UVOT/UVW2 image is used to identify UV-bright SFCs, while the GALEX FUV and NUV images provide band-separated UV photometry for the same regions \citep{2021ApJ...909..203M,2025A&A...693A.188S}. Because the UV observations do not resolve individual massive stars at the distance of NGC~5584, we interpret the UV-bright complexes as tracers of young massive-star populations associated with recent star formation, rather than as direct counts of individual OB stars. The combination of UV imaging, which traces star-forming structures on scales of several hundred parsecs, and the substantially finer-resolution NIRCam data therefore enables a disk-wide comparison among RSG overdensities, UV-selected SFCs, and near-infrared stellar light.

In this work, we quantify the relation between RSG candidates and UV-selected SFCs in NGC 5584 using JWST/NIRCam images, $Swift$/UVOT UVW2 morphology, GALEX FUV/NUV photometry, and MUSE spectroscopy. UVOT/UVW2 defines the SFC boundaries, GALEX provides separate FUV and NUV luminosities and the fiducial UV-$\beta$ extinction correction, NIRCam provides the resolved RSG-candidate catalogue and matched-aperture near-infrared luminosities, and MUSE provides an independent Balmer-decrement check for the covered subset. We focus on statistical associations and their scatter rather than interpreting spatial overlap as evidence for causal propagation or strict coevality. This paper is organized as follows. Section~2 describes the multiwavelength data, including the \textit{JWST}/NIRCam, \textit{Swift}/UVOT, GALEX, and MUSE observations. Section~3 presents the photometric measurements, RSG candidate selection, SFC definition, and UV extinction correction. Section~4 quantifies the spatial and luminosity relationships among the RSG, UV, and near-infrared tracers. Section~5 discusses the physical interpretation, including the roles of stellar age, dust extinction, spatial resolution, and aperture effects. Finally, we summarize our main results in Section~6.

\section{Data}\label{datasets}
\subsection{JWST/NIRCam data}
We utilized JWST Near-Infrared Camera (NIRCam) observations of NGC~5584 from the Cycle 1 General Observer (GO) program 1685 (PI: Adam Riess). NIRCam consists of Module A and Module B, each with a short-wavelength (SW) and a long-wavelength (LW) channel. The SW channel comprises a $2\times2$ array of four detectors, each with a field of view of $64'' \times 64''$, separated by gaps of approximately 4-5$''$, yielding a total field of view of $2.2' \times 2.2'$. The LW channel has a co-aligned $129'' \times 129''$ field of view.

\begin{table}
\centering
\caption{Basic properties of NGC~5584.}
\label{tab:ngc5584_properties}
\begin{tabular}{ccc}
\hline
Property & Value & Reference \\
\hline
Morphology & Late-type weakly barred spiral & \citet{2021MNRAS.506.1896K} \\
Distance & 23.3 Mpc & \citet{2024ApJ...962L..17R} \\
Physical scale & 113 pc arcsec$^{-1}$ & Calculated \\
Optical size & $3.13\arcmin \times 2.34\arcmin$ & \citet{2021MNRAS.506.1896K} \\
Physical size & $21.2 \times 15.9$ kpc & Calculated \\
Inclination & $42.4^\circ$ & \citet{2021MNRAS.506.1896K} \\
$\log(M_\star/M_\odot)$ & 9.95 & \citet{2021MNRAS.506.1896K} \\
$\log({\rm SFR}/M_\odot\,{\rm yr}^{-1})$ & $-0.17$ & \citet{2021MNRAS.506.1896K} \\
\hline
\end{tabular}
\end{table}

\begin{figure*}
    \centering
    \includegraphics[width=170mm]{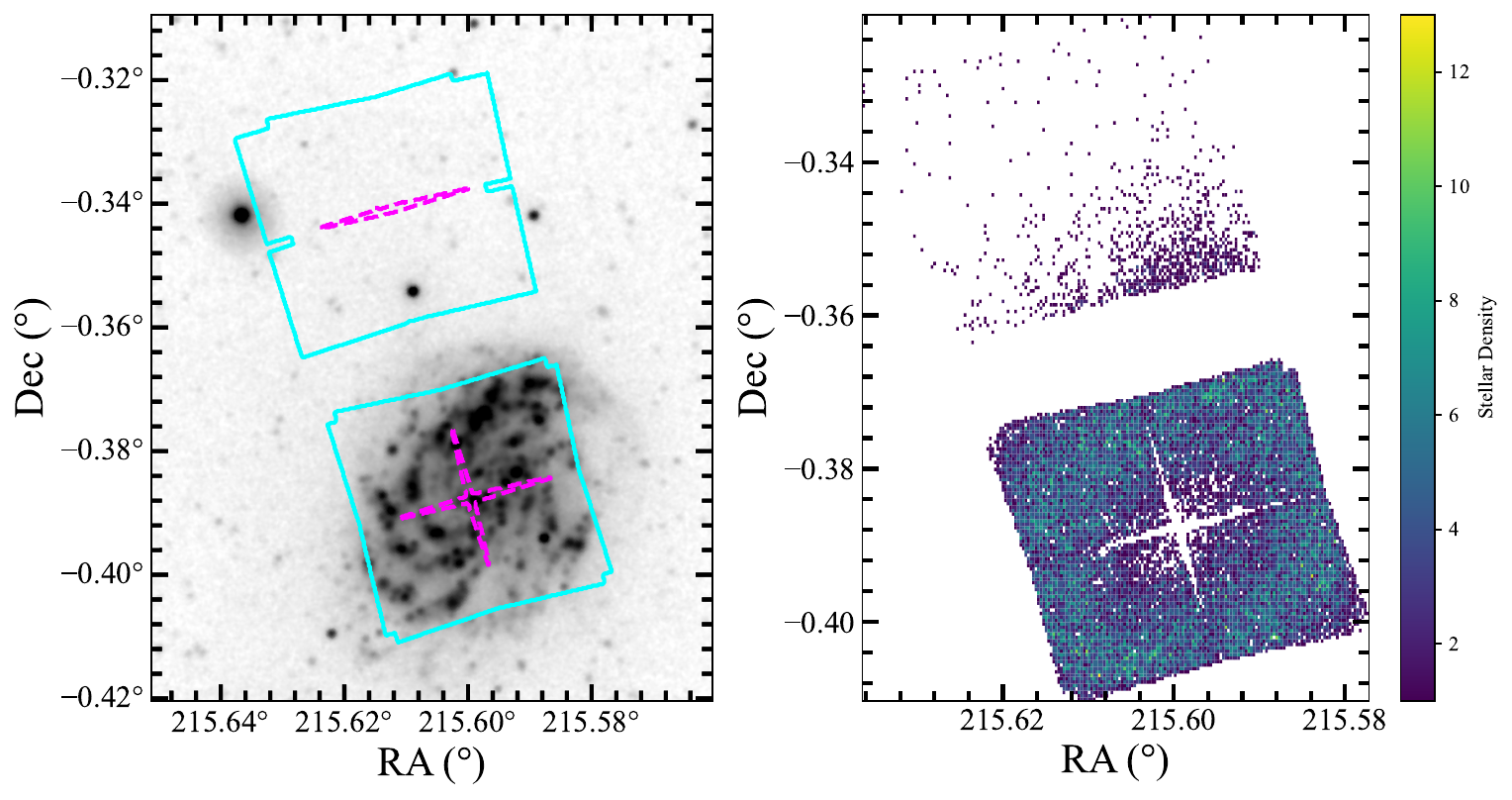}
	\caption{JWST/NIRCam observations of NGC~5584. Left: The observational footprint (solid cyan lines) is overlaid on a $Swift$/UVOT UVW2 image, with the magenta dashed lines indicating the dither pattern. Right: Spatial density map of the stars satisfying our photometric quality criteria, rendered as a 2D histogram (bins = 200). The colour represents the local stellar density, ranging from 1 (dark purple) to 13 (bright yellow).}
    \label{fig1}
\end{figure*}

NGC~5584 was observed on 2023 January 30 and 2023 February 21. Each visit included imaging in three filters: F090W (0.9\nobreak\,$\mu$m), F150W (1.5\nobreak\,$\mu$m), and F277W (2.8\nobreak\,$\mu$m), with four dithered exposures taken per filter. A slight rotation between the two epochs was employed to better cover the central gap in the SW channel. The left panel of Figure \ref{fig1} illustrates the NIRCam footprint for NGC~5584, overlaid on an image from the $Swift$/UVOT in the UVW2 (0.21\nobreak\,$\mu$m) filter. The solid cyan line denotes the outer edge of the footprint, while the dashed magenta line indicates the dither pattern. For our analysis, we utilized only the images from the first exposure of each visit. We retrieved the data from the Mikulski Archive for Space Telescopes (MAST)\footnote{\url{https://mast.stsci.edu/portal/Mashup/Clients/Mast/Portal.html}} for subsequent data reduction and analysis.

We performed PSF photometry using the NIRCam module of the software package DOLPHOT\footnote{\url{http://americano.dolphinsim.com/dolphot/}}, which is widely used for photometry of individual stars in various galaxies, including not only members of the Local Group but also those well beyond it \citep{2023ApJ...956L..18R,2024ApJ...976..177L,2024ApJS..271...47W,2025ApJ...986...54Y}. The photometry was performed on the Stage 2 *cal.fits images, by using the Stage 3 *i2d.fits images as references for image alignment and source detection. Because the SW channel has a higher angular resolution than the LW, photometry on both channels simultaneously can introduce biases into the SW-band measurements. Following the strategy recommended by \citet{2023ApJ...956L..18R} to solve this issue, photometry was taken separately on the SW images alone and on the combined SW+LW images. We then adopted the results from the SW-only run for the SW bands and those from the combined run for the LW bands. This approach ensures the photometric accuracy of the SW measurements while preserving the detection completeness in the LW channel.

We cross-matched the resulting SW-only and SW $+$ LW photometric catalogs using a 0.02$''$ radius \citep[e.g.,][]{2025ApJS..279...56W}. This process yielded an initial photometric catalog of 818,370 sources detected in all the F090W, F150W, and F277W bands. To ensure high photometric quality, we filtered the initial catalog by adopting widely-used criteria similar to those in \citet{2024ApJS..271...47W} as follows: (1) Object Type $\leq$ 2; (2) Crowding $\leq$ 0.5; (3) Sharpness$^2$ $\leq$ 0.01; (4) S/N $\geq$ 4; (5) Error Flag $=$ 2 and (6) Magnitude $<$ 30\nobreak\,mag. These criteria are applied simultaneously to all three bands (F090W, F150W, and F277W). These quality cuts resulted in a final catalog of 36,315 stars. The right panel of Figure \ref{fig1} shows the spatial distribution of these quality-selected stars.

\subsection{$Swift$/UVOT and GALEX data}
\label{sec:uv_data}

We used the ultraviolet observations for two distinct purposes. First, the \textit{Swift}/UVOT UVW2 image provides disk-wide UV morphology at a finer angular resolution than the GALEX images and is used to define the segmentation apertures of the UV-selected SFCs. The UVW2 filter has an effective wavelength of 2052.5~\AA\footnote{\url{https://svo2.cab.inta-csic.es/theory/fps/index.php}} and a passband full width at half maximum (FWHM) of 657~\AA\ \citep{2008MNRAS.383..627P}. Its image has a PSF FWHM of $2.92\arcsec$ \citep{2010MNRAS.406.1687B}, corresponding to a physical scale of approximately 330~pc at the adopted distance of 23.3~Mpc.

Second, the GALEX FUV and NUV images are used to perform quantitative UV photometry within the SFC apertures defined from the UVOT/UVW2 morphology. Although UVW2 provides the higher-resolution morphological information required for source segmentation, its broad passband does not correspond directly to either of the conventional GALEX FUV and NUV bands. We therefore do not use the UVW2 flux in the quantitative determination of the FUV and NUV luminosities or the UV continuum slope. Instead, we perform matched-aperture photometry on the GALEX FUV and NUV images using the UVOT-defined SFC apertures. The adopted effective wavelengths of the GALEX FUV and NUV bands are 1548.85 and 2303.37~\AA, respectively\footnote{\url{https://svo2.cab.inta-csic.es/theory/fps/index.php}}, and their passband FWHM values are 269 and 616~\AA, respectively\footnote{\url{https://asd.gsfc.nasa.gov/archive/galex/Documents/GALEXObserverGuide.pdf}}. The corresponding PSF FWHM values are $4.2\arcsec$ and $5.3\arcsec$ \citep{2007ApJS..173..682M}, equivalent to physical scales of approximately 474 and 599~pc. The GALEX measurements provide separate FUV and NUV luminosities for each SFC and are used to estimate the UV continuum slope required for the internal-dust extinction correction.

\subsection{MUSE spectroscopy}
\label{sec:muse_data}

\begin{figure*}
    \centering
    \includegraphics[width=85mm]{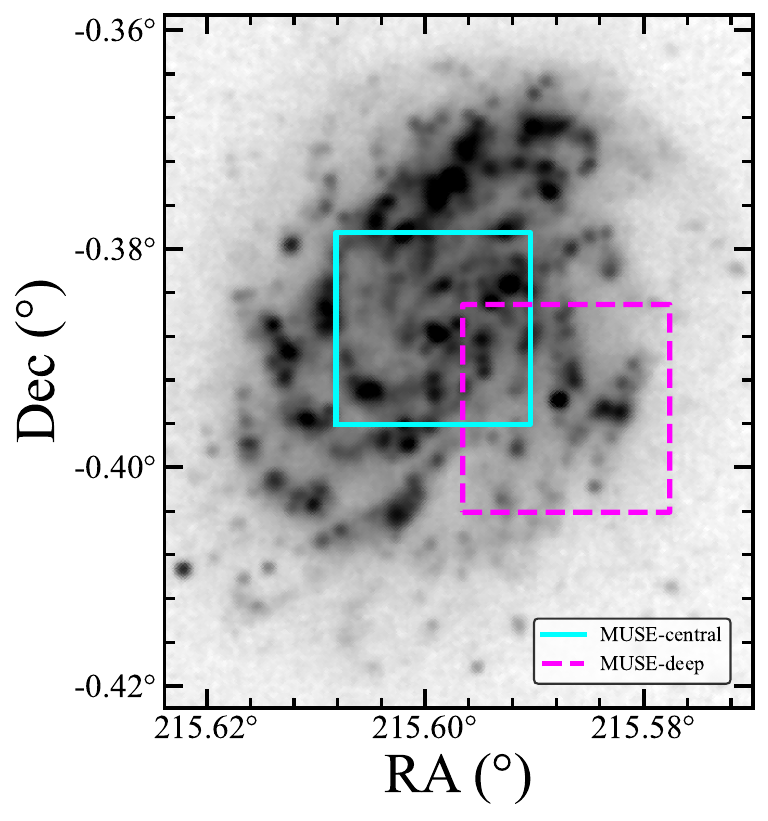}
	\caption{MUSE spectroscopic coverage adopted in this work, overlaid on the \textit{Swift}/UVOT UVW2 image of NGC~5584. The cyan solid (fov = $1.5\arcmin$) shows the MUSE-central pointing covering the central region, while the magenta dashed (fov = $1.61\arcmin$) shows the MUSE-deep pointing toward the south-eastern disk.}
    \label{fig:muse_footprint}
\end{figure*}

Archival VLT/MUSE\footnote{\url{https://archive.eso.org/scienceportal/home}} integral-field spectroscopy is available for NGC~5584. The archival data comprise one MUSE pointing covering the central region of the galaxy and several pointings covering the southeastern disk. In this work, we used two non-redundant data cubes, shown in Figure ~\ref{fig:muse_footprint}: the central MUSE pointing (solid cyan outline; FoV = $1.5\arcmin$) and the deepest southeastern pointing (magenta dashed outline; FoV = $1.61\arcmin$), hereafter referred to as MUSE-central and MUSE-deep, respectively. The other southeastern MUSE cubes cover nearly the same region as MUSE-deep but have shorter exposure times and are therefore not used, thereby avoiding duplicate, non-independent measurements of the same SFC/clump regions.

The MUSE wavelength coverage (475--935.1\,nm) includes both H$\beta$ and H$\alpha$, enabling us to measure the Balmer decrement for the subset of SFC/clump regions located within these two fields. These measurements provide an independent check on the UV extinction scale, while the UV-$\beta$ correction remains the fiducial extinction correction for the full SFC/clump sample (discussed in Section \ref{sec:balmer_check_discussion}).

\section{Sample}
\subsection{RSG Sample} \label{sec:Samples}
% \subsection{Removing the Foreground Stars}\label{subsec:rm fgd}

To identify RSG candidates in NGC~5584, we mitigated foreground Galactic star contamination within our photometric catalog by first cross-matching with Gaia DR3 astrometric solutions. This selection was complemented by NIR photometric diagnostics, specifically utilizing an F150W $-$ F277W versus F090W $-$ F150W colour-colour diagram (CCD).

We first cross-matched our catalog with Gaia DR3 to identify sources with reliable astrometric solutions and classified them as foreground stars. \citep[e.g.,][]{2019A&A...629A..91Y,2020A&A...639A.116Y,2021A&A...646A.141Y,2021ApJ...907...18R,2021A&A...647A.167Y,2025MNRAS.539.1220D,2025ApJ...988...60D}. This method definitively removed six foreground stars.

\begin{figure*}
    \centering
    \includegraphics[width=170mm]{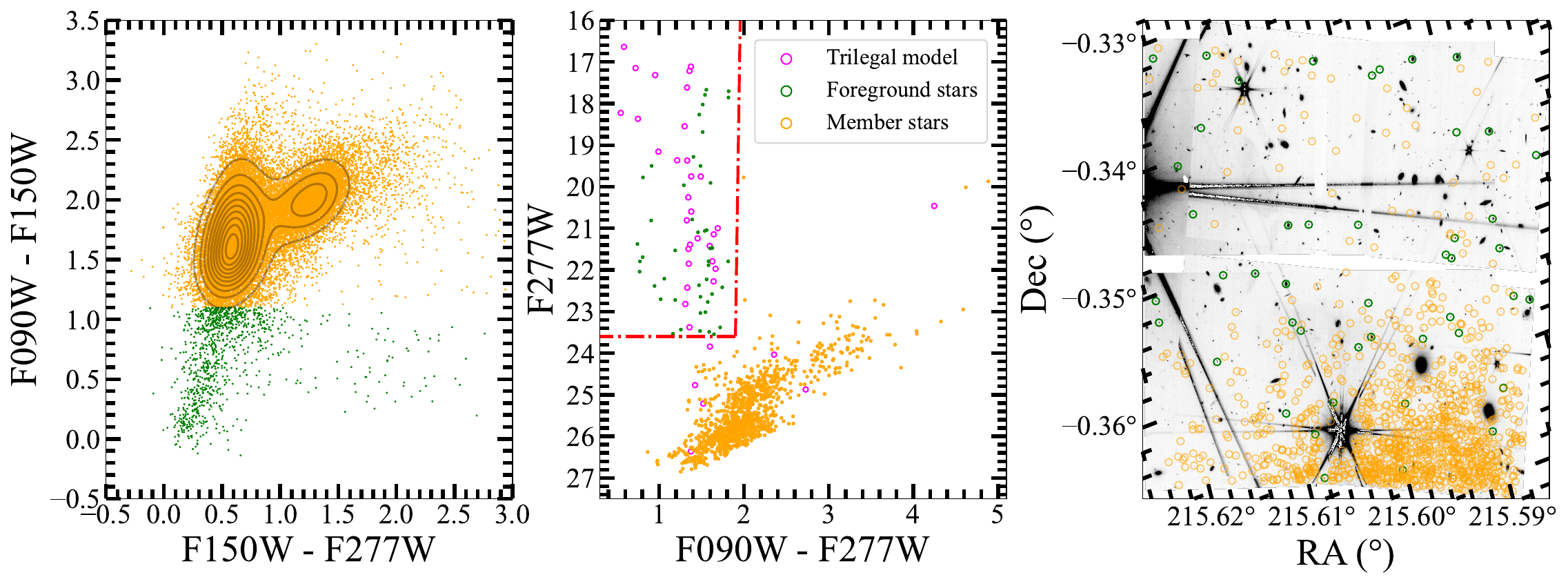}
	\caption{Separation of NGC~5584 member stars from foreground contamination. Left: The F090W$-$F150W vs. F150W$-$F277W colour-colour diagram of the NGC~5584. Green circles represent foreground stars selected using the colour criterion F090W$-$F150W $<$ 1.1, while contours delineate the stellar number density. Middle: The F277W vs. F090W$-$F277W colour-magnitude diagram of the reference region, distinguishing our sample of member stars (orange) from foreground stars (green). The red dash-dotted lines indicate the selection boundary. For comparison, simulated foreground stars from the TRILEGAL model are overplotted as magenta circles. Right: Spatial distribution of the identified member (orange) and foreground (green) stars within the reference region, overlaid on the JWST/NIRCam F150W image.}
    \label{figA1}
\end{figure*}

Next, we used photometric diagnostics. A common approach is to use a NIR CCD, which separates dwarf stars from giants and supergiants. This technique uses the fact that our F150W filter is analogous to the $H$ band (1.6\nobreak\,$\mu$m), where differences in surface gravity suppress the flux in dwarfs relative to giants and supergiants. An additional advantage of NIR CCDs is their low sensitivity to extinction effects \citep{2019ApJ...877..116W, 2024ApJ...964L...3W}. The F150W $-$ F277W versus F090W $-$ F150W CCD for our sample is shown in the left panel of Figure \ref{figA1}. The diagram clearly displays the expected bifurcation, with the stellar density distribution forming two distinct sequences: (1) a main, densely populated locus of target-field giants and supergiants, and (2) a ``bending branch'' at bluer F090W $-$ F150W colours. Following the methodology of recent studies \citep[e.g.,][]{2025ApJ...988...60D}, we identify this bluer branch, quantitatively defined as F090W $-$ F150W $<$ 1.1\nobreak\,mag, as the locus of foreground dwarf stars (green). However, due to the small (2.2$'$ $\times$ 2.2$'$) Field of View (FoV) of the JWST observations, this dwarf locus is sparsely populated and its separation from the giant branch is less clearly defined than in wide-field surveys \citep[cf. Figure 1 in][]{2025ApJ...988...60D}.

To further assess the foreground-star selection, we compared the colour-magnitude distribution in a reference field outside the main body of NGC 5584 with a TRILEGAL\footnote{\url{https://stev.oapd.inaf.it/cgi-bin/trilegal_1.6}} simulation for the same line of sight and area. The selected foreground candidates occupy the expected locus and are distributed approximately uniformly across the reference field, whereas the remaining sources are concentrated toward NGC~5584. We constructed the F277W versus F090W $-$ F277W CMD for stars in this reference field (middle panel of Figure~\ref{figA1}) and compared their distribution to a simulated foreground population from the TRILEGAL model for the same FoV. This comparison allowed us to define a selection for the foreground population as sources with F090W $-$ F277W $<$ 1.9\nobreak\,mag and F277W $<$ 23.6\nobreak\,mag (red dashed–dotted line). The right panel of Figure~\ref{figA1} provides spatial validation for this classification that the identified foreground stars (green) are distributed uniformly and mainly in the outer region as expected, while the remaining presumed member stars (orange) are clustered close to the galactic center.

\begin{figure*}
    \centering
    \includegraphics[width=170mm]{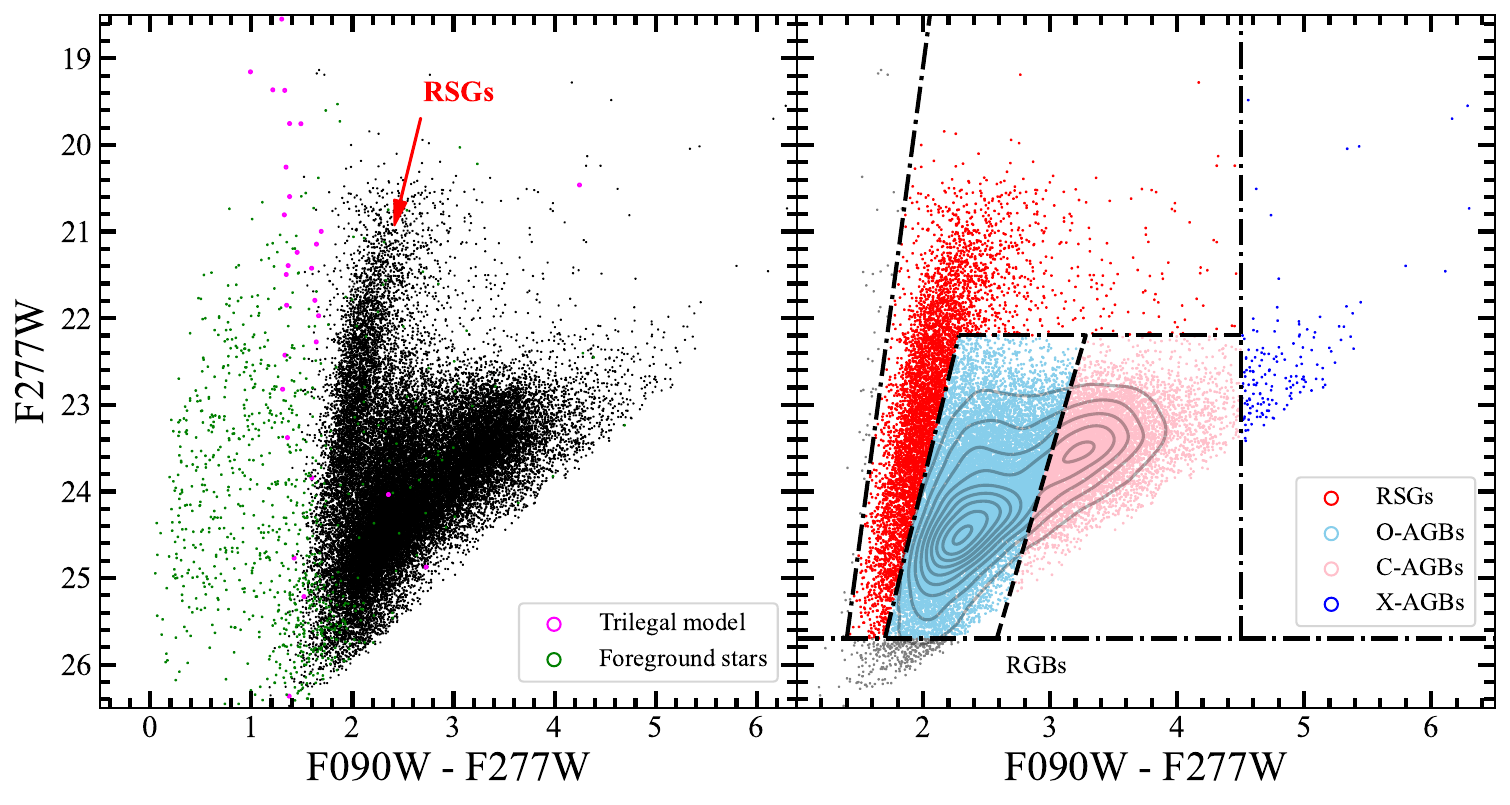}
	\caption{F277W versus F090W $-$ F277W CMDs illustrating the selection and classification of evolved stellar populations from JWST/NIRCam photometry.
Left: The CMD shows the host galaxy’s stellar population (black points), where the RSG branch is clearly visible, as indicated by the red arrows. For comparison, predicted foreground Milky Way stars (green) and a TRILEGAL stellar population model (magenta) are overlaid.
Right: Classification of the member stars into RSGs (red), O-AGBs (sky blue), C-AGBs (pink), X-AGBs (blue) and RGBs (gray) sub-populations. Density contours are provided for the O-AGB and C-AGB regions to illustrate the population density. The horizontal dash-dotted line at the bottom indicates the magnitude of the TRGB, which serves as the lower luminosity limit for the RSGs and AGBs selection.}
    \label{fig2}
\end{figure*}

The left panel of Figure \ref{fig2} demonstrates the overall effectiveness of our approach. The CMD shows that our photometrically-selected foreground stars (green) align well with the population predicted by Trilegal (magenta). It is evident that while some foreground contamination exists, particularly at the faint end of RSG, it is minimal for the most luminous RSGs. This agreement supports our foreground removal process.

After removing contamination from foreground stars, the RSG branch is clearly visible in the F277W versus F090W$-$F277W CMD (see the left panel of Figure \ref{fig2}, indicated by red arrows.) Following standard practice, we set the faint luminosity limit for RSGs at the Tip of the Red Giant Branch (TRGB) \citep[e.g.,][]{2021ApJ...907...18R,2021ApJ...923..232R,2025ApJ...979..208L,2025ApJ...988...60D}. Adopting an absolute TRGB magnitude of $M_{\mathrm{F277W,TRGB}} = -6.14$\nobreak\,mag \citep{2024ApJ...975..195N} and a distance modulus of $\mu_{0} = 31.838$\nobreak\,mag for NGC~5584 \citep{2024ApJ...962L..17R}, we calculated an apparent magnitude of $m_{\mathrm{F277W,TRGB}} = 25.698$\nobreak\,mag. Guided by the stellar density contours, we define the full selection region with the following boundaries:
% \begin{equation}
% 	\text{left\ boundary}: \mathrm{F277W - 25.698 = -11(F090W-F277W-1.4),}
% 	\label{eq1}
% \end{equation}
% \begin{equation}
% 	\text{lower\ right\ boundary}: \mathrm{F277W - 25.698 = -6(F090W-F277W-1.7),}
% \end{equation}
% \begin{equation}
% 	\text{upper\ right\ boundary}: \mathrm{F090W - F277W = 4.5,}
% \end{equation}
% \begin{equation}
% 	\text{middle\ horizontal\ boundary}: \mathrm{F277W = 22.198,}
% \end{equation}
% \begin{equation}
% 	\text{Lower\ horizontal\ boundary}: \mathrm{F277W = 25.698.}
% 	\label{eq5}
% \end{equation}
% Guided by the stellar density contours, we define the selection region with the following boundaries (see also Fig. 1):
\begin{small} 
\begin{equation}
\label{eq:selection}
\left\{
\begin{aligned}
    & \text{Left: } \mathrm{F277W - 25.698 = -11(F090W-F277W-1.4)} \\
    & \text{Lower right: } \mathrm{F277W - 25.698 = -6(F090W-F277W-1.7)} \\
    & \text{Upper right: } \mathrm{F090W - F277W = 4.5} \\
    & \text{Middle horizontal: } \mathrm{F277W = 22.198} \\
    & \text{Lower horizontal: } \mathrm{F277W = 25.698}
\end{aligned}
\right.
\end{equation}
\end{small}

The right panel of Figure \ref{fig2} shows the CMD for NGC~5584 members, with our selection boundaries (black dashed–dotted lines) and the resulting RSG sample (red). The selected RSGs form two distinct features: a prominent, slanted sequence at F090W $-$ F277W $\approx$ 1.5--2.0\nobreak\,mag, and a more luminous horizontal branch. We classify all stars in this bright horizontal region as RSGs, since Asymptotic Giant Branch (AGB) stars are not expected to reach such high luminosities (at least 3.5\nobreak\,mag brighter than the TRGB). The redder colours of these horizontal branch RSGs likely indicate the presence of significant circumstellar dust. Using similar density-based criteria, we also identify oxygen-rich AGB (O-AGB), carbon-rich AGB (C-AGB) and extreme AGB (X-AGB) stars. Our final catalog consists of 5,310 RSGs (the candidate sample table, containing coordinates and photometry in three NIR bands, is available in Table \ref{t1}), 20,242 O-AGB, 8,202 C-AGB, and 157 X-AGB stars. To validate this selection, we examined the spatial distribution of the RSG candidates. As shown in Figure \ref{figB1}, the identified RSGs clearly trace the galaxy’s spiral arms and align with UV-detected star-forming regions. This spatial correlation confirms that our photometric cuts effectively isolate a young stellar population associated with the disk. Furthermore, we find that relaxing the crowding criterion to $\le 1.0$ increases the RSG statistics by 74.2\% without introducing substantial contamination, providing a more complete sample for statistical analysis (see Section \ref{appendix:B} for a detailed discussion on completeness).

\begin{table*}
\caption{\label{t1}Catalog of RSG candidates identified in NGC~5584. A full version of this table is available as supplementary material.}
\centering
\begin{tabular}{cccccc}
\hline\hline
No.&RA (J2000)&Dec (J2000)&F090W&F150W&F277W\\
& (deg) & (deg) & (mag) & (mag) & (mag)\\
\hline
NGC~5584-1     &215.5779021     &-0.3978255 &26.104	&24.704&	24.182\\
NGC~5584-2	&215.5782793	&-0.3969477 	&25.924&	24.355&	23.948\\
...&...&...&...&...&...\\
NGC~5584-5309	&215.6197458	&-0.3766912 	&27.338	&25.820&	25.647\\
NGC~5584-5310	&215.6197738 &-0.3831227	&26.843&	25.534&	25.047\\
\hline
\multicolumn{6}{l}{The columns list the target ID, J2000 coordinates, and JWST/NIRCam magnitudes.}
\end{tabular}
\end{table*}
%%%%%%%%%%%%%%%%%%%%%%%%%%%%%%%%%%%%%%%%%%%%%%%%%%%%%%%%%%%%%%

\subsection{UV-selected SFCs/clumps}

A complete census of individual OB stars cannot be constructed from the available NIRCam near-infrared photometry at the distance of NGC~5584. We nevertheless identified 722 luminous NIR OB candidates using the criterion F090W $-$ F150W $<0$. Their coordinates and photometry in the three NIRCam bands are provided in Table~\ref{t2}. Because this sample is deliberately incomplete, we used it only as a diagnostic of whether UV-bright regions contain detected OB sources. Accordingly, statistics based on NIR OB candidate counts are not interpreted as complete OB star fractions or as direct estimates of the intrinsic star-formation rate. Instead, the primary tracer of young massive-star formation is the set of UV-selected SFCs defined from the \textit{Swift}/UVOT UVW2 morphology.

This choice is motivated by previous studies that employed UV-bright SFCs as tracers of young massive-star populations \citep[e.g.,][]{2021ApJ...909..203M,2025A&A...693A.188S}. Our SFC-identification procedure, however, is not designed to reproduce the dendrogram-based Astrodendro method adopted in those studies. Instead, we used an independent image-segmentation approach implemented with \texttt{photutils} to define a uniform set of UV-selected SFC apertures for the subsequent matched-aperture photometry and spatial-association analyses.

We identified these SFCs/clumps using the \texttt{photutils} package in Python. First, we applied a 3$\sigma$ threshold above the median image flux to create an initial segmentation mask. We then segmented the image, requiring an initial clump size of at least 50 pixels, before deblending sources using a multi-thresholding approach (\texttt{nlevels}=128, \texttt{contrast}=0.001). The UVOT/UVW2 FITS image used in this analysis was generated and astrometrically reprojected through Aladin\footnote{\url{https://aladin.cds.unistra.fr/aladin.gml}}. Its WCS pixel scale is approximately 0.306 arcsec pixel$^{-1}$, which describes the pixel sampling of this reprojected image rather than the native detector pixel scale or the effective angular resolution of UVOT. On this image grid, 50 pixels correspond to an area of approximately 4.67 arcsec$^{2}$, equivalent to a circular radius of 1.22 $\arcsec$ (approximately 138 pc at 23.3 Mpc) or a diameter of 2.44 $\arcsec$ (approximately 275 pc).

We selected the 50-pixel threshold through iterative manual testing and visual inspection of the resulting segmentation maps. Smaller thresholds produced excessive fragmentation and numerous small islands, whereas larger thresholds tended to merge visually distinct UV-bright structures into overly extended complexes. We therefore adopted 50 pixels as a practical compromise for the initial detection stage, rather than as a measure of the effective spatial resolution or a strict minimum physical size for the final clumps.
 Finally, the boundaries of each resulting clump were refined by removing pixels below a local 1$\sigma$ threshold, ensuring that they trace the densest UV emission. Because deblending and subsequent boundary refinement can reduce the area of an individual region, the final clumps are not required to contain at least 50 pixels. Their final equivalent radii range from approximately 70 to 630 pc, with a median of approximately 265 pc. As shown in panel (a) of Figure~\ref{fig3}, this process yielded a total of 106 UV-bright clumps, which are outlined by magenta and green contours and numbered spatially.

\begin{table*}
\centering
\caption{Example rows from the SFC/clump catalogue. The complete catalogue is available in machine-readable form.}
\label{tab:sfc_clump_catalog}
\scriptsize
\setlength{\tabcolsep}{2.0pt}
\begin{tabular}{ccccccccccccccccc}
\hline
ID & RA (J2000) & Dec (J2000) & Area & $R_{\rm eq}$ & $\log L_{\rm FUV,obs}$ & $\log L_{\rm NUV,obs}$ & $A_{\rm FUV}$ & $A_{\rm NUV}$ & $\log L_{\rm FUV,corr}$ & $\log L_{\rm NUV,corr}$ & $\log L_{\rm F090W}$ & $\log L_{\rm F150W}$ & $\log L_{\rm F277W}$ & $N_{\rm OB}$ & $N_{\rm RSG}$ & Flag \\
 & (deg) & (deg) & (arcsec$^2$) & (pc) & (erg s$^{-1}$) & (erg s$^{-1}$) & (mag) & (mag) & (erg s$^{-1}$) & (erg s$^{-1}$) & (erg s$^{-1}$) & (erg s$^{-1}$) & (erg s$^{-1}$) &  & & \\
\hline
1 & 215.62210 & -0.40923 & 4.58 & 136.3 & 38.228 & 38.398 & 2.498 & 2.020 & 39.338 & 39.328 & -- & -- & -- & 0 & 0 & 1 \\
2 & 215.59302 & -0.40285 & 4.02 & 127.7 & 39.217 & 39.129 & 1.039 & 0.840 & 39.744 & 39.587 & 39.968 & 39.821 & 39.216 & 0 & 1 &  \\
3 & 215.61607 & -0.39786 & 15.04 & 247.2 & 40.027 & 39.959 & 1.156 & 0.935 & 40.601 & 40.455 & -- & -- & -- & 1 & 3 & 1 \\
4 & 215.59165 & -0.39817 & 8.97 & 190.8 & 39.796 & 39.650 & 0.709 & 0.573 & 40.192 & 40.001 & 40.381 & 40.259 & 39.671 & 2 & 15 &  \\
\multicolumn{17}{c}{\ldots} \\
103 & 215.59008 & -0.36915 & 67.07 & 521.9 & 41.100 & 41.018 & 1.072 & 0.867 & 41.640 & 41.486 & 41.315 & 41.174 & 40.635 & 9 & 86 &  \\
104 & 215.58807 & -0.36891 & 23.26 & 307.4 & 40.556 & 40.471 & 1.055 & 0.853 & 41.090 & 40.934 & 40.822 & 40.686 & 40.114 & 1 & 38 &  \\
105 & 215.59859 & -0.36826 & 3.92 & 126.2 & 39.389 & 39.246 & 0.729 & 0.589 & 39.793 & 39.604 & -- & -- & -- & 0 & 0 & 1 \\
106 & 215.59004 & -0.36588 & 8.31 & 183.8 & 39.704 & 39.543 & 0.628 & 0.508 & 40.067 & 39.868 & -- & -- & -- & 1 & 0 & 1 \\
\hline
\end{tabular}
\vspace{2pt}\par\noindent\footnotesize{All luminosities are expressed as
$\log(\nu L_\nu)$ in units of erg s$^{-1}$.
The listed $A_{\rm FUV}$ and $A_{\rm NUV}$ values are the internal
extinction corrections derived from UV-$\beta$.
The corrected FUV and NUV luminosities include both Galactic foreground
extinction and internal extinction.
Rows with $\mathrm{Flag}=1$ are excluded from the UV--JWST
luminosity-relation analysis because they lie outside the JWST/NIRCam
footprint or near the mosaic edge.}
\end{table*}

\begin{table*}
\caption{\label{t2}Catalog of OB star candidates identified in NGC 5584. A full version of this table is available as supplementary material.}
\centering
\begin{tabular}{cccccc}
\hline\hline
No.&RA (J2000)&Dec (J2000)&F090W&F150W&F277W\\
& (deg) & (deg) & (mag) & (mag) & (mag)\\
\hline
NGC~5584-1	&215.5796701 	&-0.3908494	&26.062&	26.098&	99.999\\
NGC~5584-2	&215.5803771	&-0.3913883	&25.855&	25.991&	26.496\\
...&...&...&...&...&...\\
NGC~5584-721	&215.6177980 	&-0.3807350 	&26.769	&27.230	&25.943\\
NGC~5584-722	&215.6180085 	&-0.3846651 	&25.914	&26.030	&25.921\\
\hline
\multicolumn{6}{l}{Columns are the same as in Table~\ref{t1}. A magnitude of 99.999 indicates a non-detection.}\\
\end{tabular}
\end{table*}
\section{Association with OB stars} \label{sec:Association}

\subsection{Spatial Correlation Analysis}\label{sec:Spatial_Analysis}

\begin{figure*}
    \centering
    \includegraphics[width=170mm]{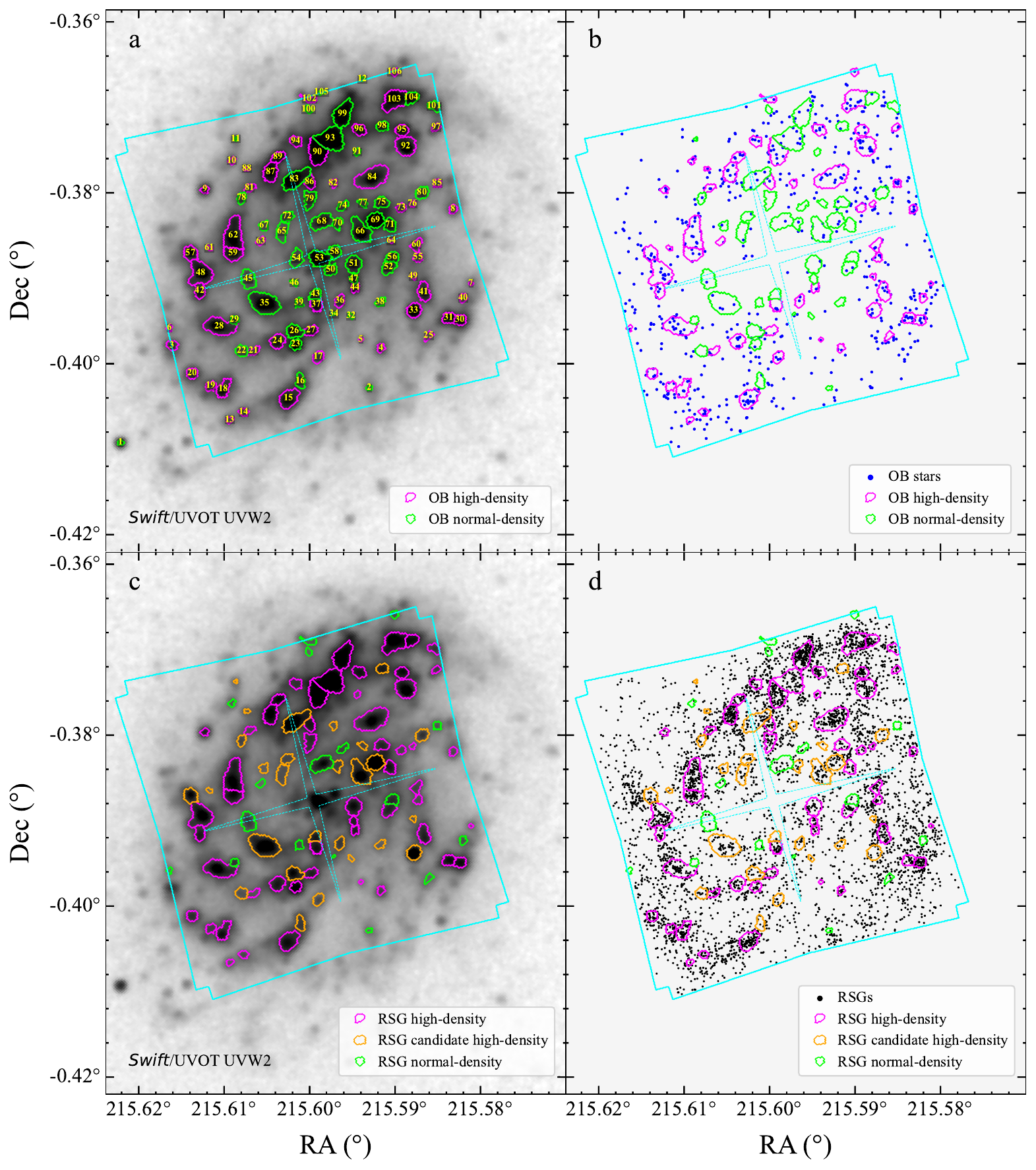}
	\caption{Spatial clustering of OB stars and RSGs in NGC~5584. (a) Distribution of OB star clumps overlaid on the Swift/UVOT UVW2 image. Magenta and green contours delineate high- and normal-density stellar clumps, respectively. Yellow numbers indicate the identification indices of these clumps. (b) Individual OB stars (blue dots) identified with JWST/NIRCam. The contours are the same as in panel (a). (c) RSG population and identified clumps overlaid on the Swift/UVOT UVW2 image. Magenta, orange, and green contours represent high-density, candidate high-density, and normal-density RSG clumps, respectively. (d) Spatial distribution of individual RSGs (black dots) identified from JWST/NIRCam photometry. The close spatial correlation between RSG clumps and UV-bright regions in panel (c) suggests a strong association with recent star formation episodes. In all panels, the cyan lines delineate the JWST/NIRCam observational footprint.}
    \label{fig3}
\end{figure*}

To validate our UV clumps as reliable OB association proxies, we quantified the density of JWST-identified OB stars within them. We calculated the average surface density of OB candidates across the field ($\rho_{\mathrm{avg}}^{\mathrm{OB}}$) and the local density within each UV clump ($\rho_{i, \mathrm{clump}}^{\mathrm{OB}}$). We then classified clumps as high-density if $\rho_{i, \mathrm{clump}}^{\mathrm{OB}} > 2 \rho_{\mathrm{avg}}^{\mathrm{OB}}$ and normal-density otherwise. Before analysis, we excluded 10 clumps: five located outside the JWST photometric footprint and five near the crowded galactic center where photometry is unreliable. The fact that a majority (56/96, 58\%) of the remaining clumps are classified as high-density (magenta), combined with the clear visual correspondence in Figure \ref{fig3} (panel a and b), confirms that our UV-bright clumps are excellent proxies for OB associations. 

Having validated the UV clumps as tracers of young massive star formation, we then measured the spatial density of RSGs within these same regions. As shown in Figure \ref{figB1}, our identified RSGs (red dots) are mainly located along the spiral arms of NGC~5584. This spatial correlation with active star-forming regions is expected for a young massive stellar population. To quantify this, we followed a similar methodology and classified the clumps based on their local RSG density ($\rho_{i,\ \mathrm{clump}}^{\mathrm{RSG}}$) relative to the field average ($\rho_{\mathrm{avg}}^{\mathrm{RSG}}$) into three categories: high-density ($\rho_{i, \mathrm{clump}}^{\mathrm{RSG}} > 2 \rho_{\mathrm{avg}}^{\mathrm{RSG}}$, magenta), candidate high-density ($\rho_{\mathrm{avg}}^{\mathrm{RSG}} < \rho_{i, \mathrm{clump}}^{\mathrm{RSG}} \le 2 \rho_{\mathrm{avg}}^{\mathrm{RSG}}$, orange), and normal-density ($\rho_{i, \mathrm{clump}}^{\mathrm{RSG}} < \rho_{\mathrm{avg}}^{\mathrm{RSG}}$, green), as shown in the panel (c) and (d) of Figure \ref{fig3}.

Our analysis reveals that the vast majority of these young, UV-defined regions are also sites of significant RSG overdensity. In total, 82\% of the clumps are classified as either high-density (52 clumps, 54\%) or candidate high-density (27 clumps, 28\%) in RSGs. This result indicates a strong galaxy-scale statistical association between RSG overdensities and UV-selected SFCs in NGC~5584. 

It is important to note that the RSGs distribution is not expected to show a one-to-one correspondence with the current UV emission. Panel d of Figure~\ref{fig3} shows that many individual RSG candidates are located outside the UV-bright SFC/clump boundaries. This is not unexpected, because RSGs and UV emission trace different phases of recent massive-star formation rather than identical instantaneous structures. UV-bright regions are dominated by younger massive stars, while RSGs appear at a later evolutionary stage and can be displaced from the current UV morphology by stellar drift, association dispersal, dust extinction, and the fading of UV emission. We therefore do not assume a one-to-one correspondence between individual RSGs and UV-bright clumps; instead, the analysis focuses on the statistical association between UV-selected SFCs and RSG overdensities on the scale of SFCs.

\subsection{FUV/NUV luminosities versus JWST/NIRCam luminosities}
\label{sec:uv_jwst_results}

We compare the GALEX FUV and NUV luminosities with the \textit{JWST}/NIRCam luminosities measured within the same UV-selected SFC/clump apertures. Similar matched-aperture comparisons between UV,
optical, infrared, and nebular tracers have been used to quantify the multi-wavelength properties of SFCs in nearby galaxies \citep[e.g.,][]{2024ApJS..271....2H,2025ApJ...987...11A}. Here, this
comparison is used to quantify the relation between the young massive-star populations traced by the UV emission and the near-infrared stellar light detected by NIRCam. We do not adopt an OB/(OB+RSG) fraction as the primary statistic because the NIR-selected OB-candidate catalogue is incomplete and spatially non-uniform.

\subsubsection{Matched-aperture UV and JWST photometry of SFCs}
\label{sec:matched_aperture_photometry}

The SFC regions used throughout this work are defined from the \textit{Swift}/UVOT UVW2 image. The resulting segmentation map contains 106 labelled regions and is defined in the UVOT WCS frame. To compare the UV and near-infrared emission consistently, we project the same segmentation map onto the GALEX FUV/NUV images and the \textit{JWST}/NIRCam images, and measure integrated fluxes within the same SFC/clump apertures.

\begin{figure*}
    \centering
    \includegraphics[width=170mm]{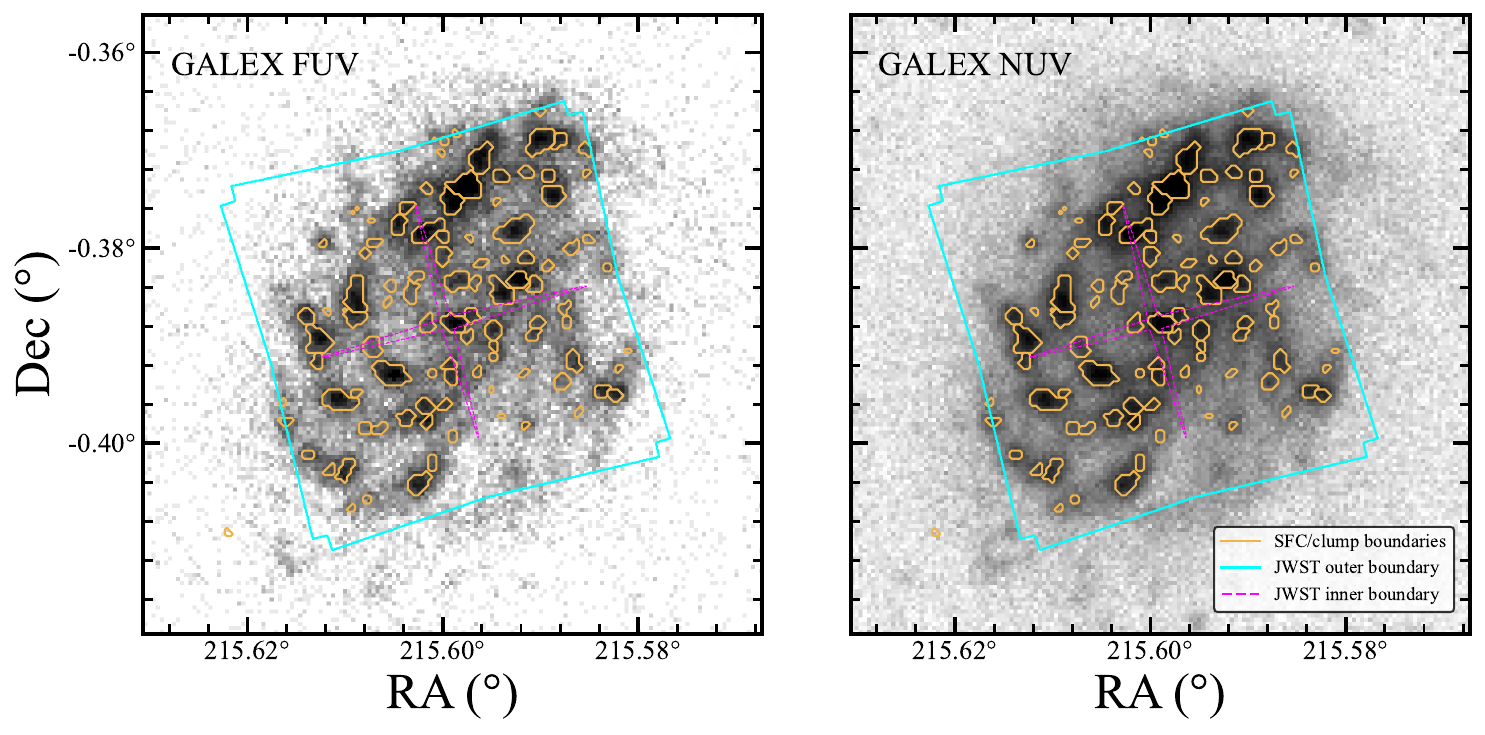}
    \caption{GALEX ultraviolet images of NGC~5584 with the SFC/clump apertures and JWST/NIRCam coverage overlaid. The left and right panels show the GALEX FUV and NUV images, respectively. Orange contours mark the UV-selected
    SFC/clump boundaries, which are defined from the \textit{Swift}/UVOT UVW2 morphology and projected onto the GALEX images for FUV and NUV photometry. The cyan solid contour shows the outer JWST/NIRCam footprint, while the magenta dashed contour indicates the inner boundary of the JWST
    coverage.}
    \label{fig:galex_clumps_jwst}
\end{figure*}

For the GALEX measurements, we used the FUV and NUV intensity images, together with the corresponding pipeline sky-background maps, exposure maps, and flag maps. The UVOT segmentation map is projected onto each GALEX image using the UVOT and GALEX WCS solutions. Because the GALEX pixels are larger than the pixels in the UVOT segmentation map, we apply a $5\times5$ subpixel sampling scheme to each GALEX pixel to estimate its fractional overlap with each SFC/clump mask. For each clump in each band, we measure the total count rate, the pipeline sky-subtracted count rate, the valid projected area, the exposure-weighted mean exposure time, and the fraction of the area affected by nonzero GALEX flags. The nonzero flag fraction is retained as a quality
diagnostic; by default, flagged pixels are not excluded from the fiducial photometry because the coarse sampling of the GALEX flag image could otherwise remove a large fraction of the area within small clumps.

The GALEX count rates are converted to AB magnitudes and physical fluxes using the standard GALEX calibration constants. We adopt effective wavelengths of 1548.85 and 2303.37~\AA\ for the FUV and NUV bands\footnote{\url{http://svo2.cab.inta-csic.es/theory/fps/}}, AB zero-points of 18.82 and 20.08, and count-rate-to-flux-density conversion factors of $1.40\times10^{-15}$ and $2.06\times10^{-16}$ erg s$^{-1}$ cm$^{-2}$ \AA$^{-1}$ per count s$^{-1}$ for FUV and NUV, respectively \citep{2007ApJS..173..682M}. The luminosities are reported as $\nu L_\nu$ using the adopted distance of 23.3~Mpc.

For the \textit{JWST}/NIRCam measurements, the same UVOT-defined segmentation map is projected onto the level-3 i2d images for F090W, F150W, and F277W. The NIRCam pixel scale is much finer than the characteristic angular sizes of the UVOT-defined SFCs; therefore, we perform the fiducial aperture integration directly on the native NIRCam pixel grid. The level-3 i2d images are calibrated in units of MJy sr$^{-1}$; we convert the summed surface brightness in each projected clump to an integrated flux density using the pixel solid angle recorded in the FITS header. We then convert the flux densities to $\nu F_\nu$ and $\nu L_\nu$ using effective wavelengths of 8987.50, 14874.09, and 27410.01~\AA\ for F090W, F150W, and F277W\footnote{\url{https://svo2.cab.inta-csic.es/theory/fps/index.php}}, respectively.

\subsubsection{Internal extinction correction}
\label{sec:uv_beta_correction}

We used the GALEX FUV and NUV measurements to estimate the UV extinction correction for all SFCs. We adopt the UV-$\beta$ slope method \citep[e.g.,][]{1994ApJ...429..582C,2025ApJ...987...11A} as the fiducial internal extinction correction for the UV--JWST comparison, because it is available for the full UV-selected SFC/clump sample. The Balmer-decrement
method is used only as an independent check for the subset of regions covered by the MUSE observations.

The correction is applied in two steps. First, we correct the observed GALEX photometry for Milky Way foreground extinction. We adopt $A_V=0.107$ mag\footnote{\url{https://ned.ipac.caltech.edu/}} and $R_V=3.1$, and use the CCM89 Galactic extinction curve \citep{1989ApJ...345..245C,2011ApJ...737..103S,2019ApJ...877..116W}. This gives foreground corrections of $A_{\rm FUV}=0.279$ mag and $A_{\rm NUV}=0.305$ mag for the GALEX bands. The UV continuum slope is then computed from the foreground-corrected GALEX FUV and NUV AB magnitudes as
\begin{equation}
\beta =
\frac{m_{\rm FUV}-m_{\rm NUV}}
{-2.5\log_{10}(\lambda_{\rm FUV}/\lambda_{\rm NUV})}
-2 ,
\end{equation}
where $\lambda_{\rm FUV}=1548.85$~\AA\ and
$\lambda_{\rm NUV}=2303.37$~\AA. We then infer the UV-slope-based colour-excess parameter,

\begin{equation}
E(B-V)_{\beta}
=
\max\left[
0,\,
\frac{\beta+2.616}{4.594}
\right],
\label{eq:ebv_beta}
\end{equation}

following the calibration of \citet{2018ApJ...853...56R} as implemented by \citet{2025ApJ...987...11A}. Following the empirical prescription adopted by \citet{2025ApJ...987...11A}, we calculated the internal UV extinction as
\begin{equation}
A_\lambda = 0.44\,E(B-V)\,k_{\rm Calzetti}(\lambda),
\end{equation}
where $k_{\rm Calzetti}(\lambda)$ is the extinction curve of \citet{2000ApJ...533..682C}. Evaluating this curve at the GALEX effective
wavelengths gives the band-dependent internal extinctions used to correct the FUV and NUV luminosities.

For the full sample of 106 SFC/clump regions, the median foreground-corrected UV slope is $\beta=-1.400$. The inferred internal extinction has median values of $A_{\rm FUV}^{\rm int}=1.182$ mag and
$A_{\rm NUV}^{\rm int}=0.956$ mag, with ranges of $0.475$--$2.498$ mag and $0.384$--$2.020$ mag, respectively. These UV-$\beta$-corrected luminosities are used as the fiducial UV luminosities in the following comparison with the NIRCam measurements.

\begin{figure*}
    \centering
    \includegraphics[width=170mm]{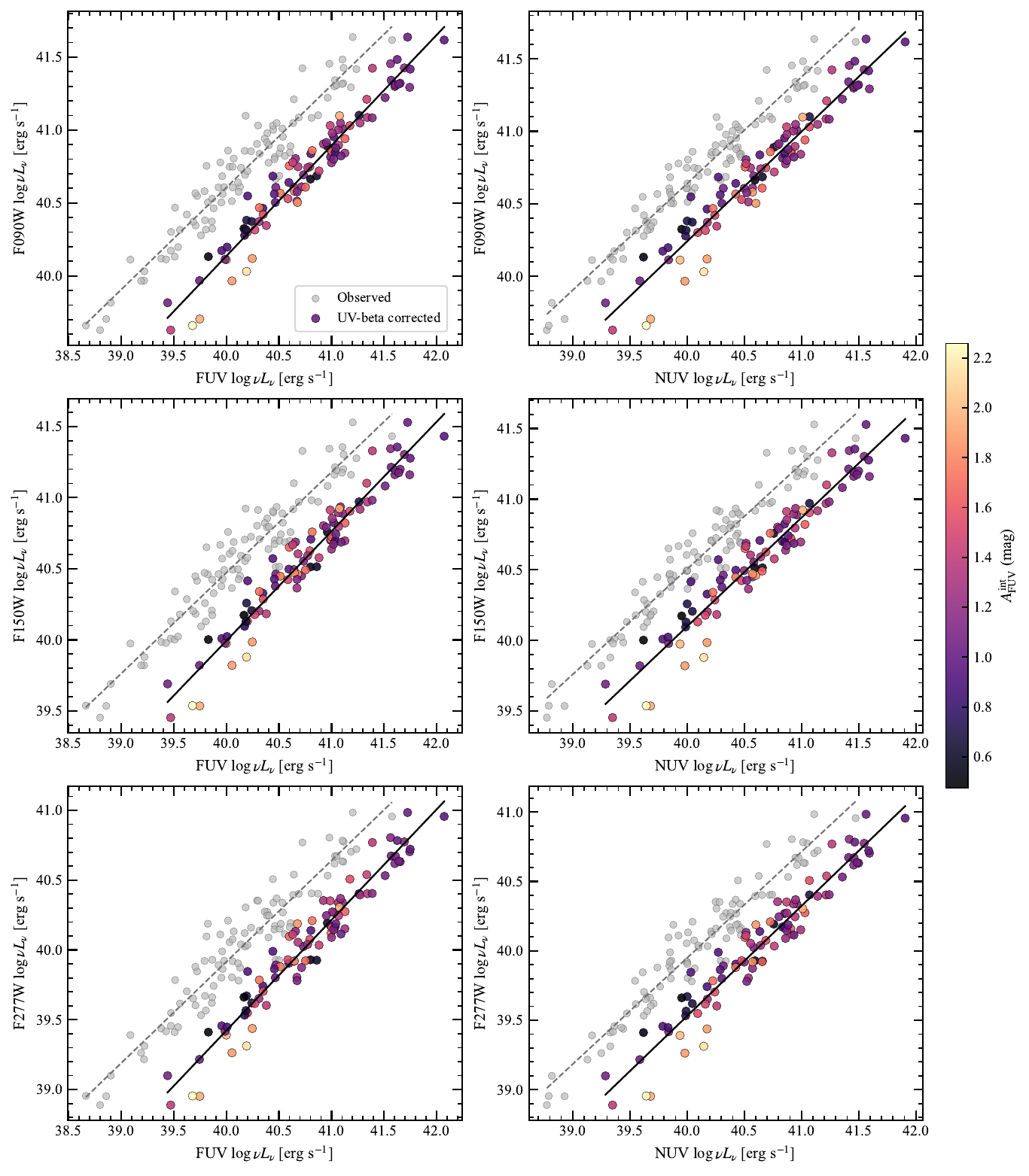}
    \caption{Comparison between GALEX UV luminosities and JWST/NIRCam near-infrared luminosities for the UV-selected SFC/clump sample. The left and right columns show the GALEX FUV and NUV bands, respectively, while the three rows correspond to the NIRCam F090W, F150W, and F277W bands. Grey points show the observed UV luminosities, and coloured points show the luminosities after the UV-$\beta$ internal extinction correction. The colour scale indicates the internal FUV extinction, $A_{\rm FUV}^{\rm int}$. Dashed grey and solid black lines show linear fits to the observed and UV-$\beta$-corrected relations, respectively. clumps with limited JWST/NIRCam coverage, either outside the footprint or close to the mosaic edge, are excluded from the fitted sample and are not shown. The UV-extinction correction shifts the UV luminosities to higher values, while the UV--near-infrared correlations remain strong, indicating that the UV-selected SFCs and the near-infrared stellar light measured with JWST/NIRCam trace closely related recent star-forming structures.}
    \label{fig:uv_jwst_relation}
\end{figure*}

Figure~\ref{fig:uv_jwst_relation} compares the GALEX FUV and NUV luminosities with the JWST/NIRCam F090W, F150W, and F277W luminosities for the UV-selected SFC/clump sample. Grey points show the observed UV luminosities, whereas the coloured points show the UV-$\beta$ extinction-corrected luminosities. For the quantitative fits, we additionally excluded clumps 6, 102, and 106 because of their severely incomplete JWST/NIRCam coverage, leaving $N=93$ clumps.

The extinction-corrected UV and near-infrared luminosities are strongly correlated across the SFC/clump sample. For the FUV relations, the Pearson
correlation coefficients are $r=$ 0.971, 0.968, and 0.968 for F090W, F150W, and F277W, respectively. The NUV relations are similarly strong, with $r=$ 0.964, 0.962, and 0.963 for F090W, F150W, and F277W,
respectively.

These correlations demonstrate that the UV-selected SFCs and the NIRCam near-infrared stellar structures are closely related on the scale of individual SFC/clump regions. At the same time, the relations are not strictly one-to-one: the best-fit slopes are shallower than unity, and a finite scatter remains after the UV-$\beta$ extinction correction. This behaviour is expected because the FUV, NUV, and near-infrared bands are sensitive to different stellar populations and respond differently to dust extinction, stellar age, aperture coverage, and
local star-formation history. The FUV band is most sensitive to the youngest massive stars, the NUV band traces a somewhat broader young stellar population,
and the NIRCam bands include contributions from both young and mainly more evolved stellar components.

\section{Discussion} \label{sec:Discussions}
\subsection{RSG Sample Completeness and Spatial Distribution}\label{appendix:B}
\begin{figure*}
    \centering
    \includegraphics[width=170mm]{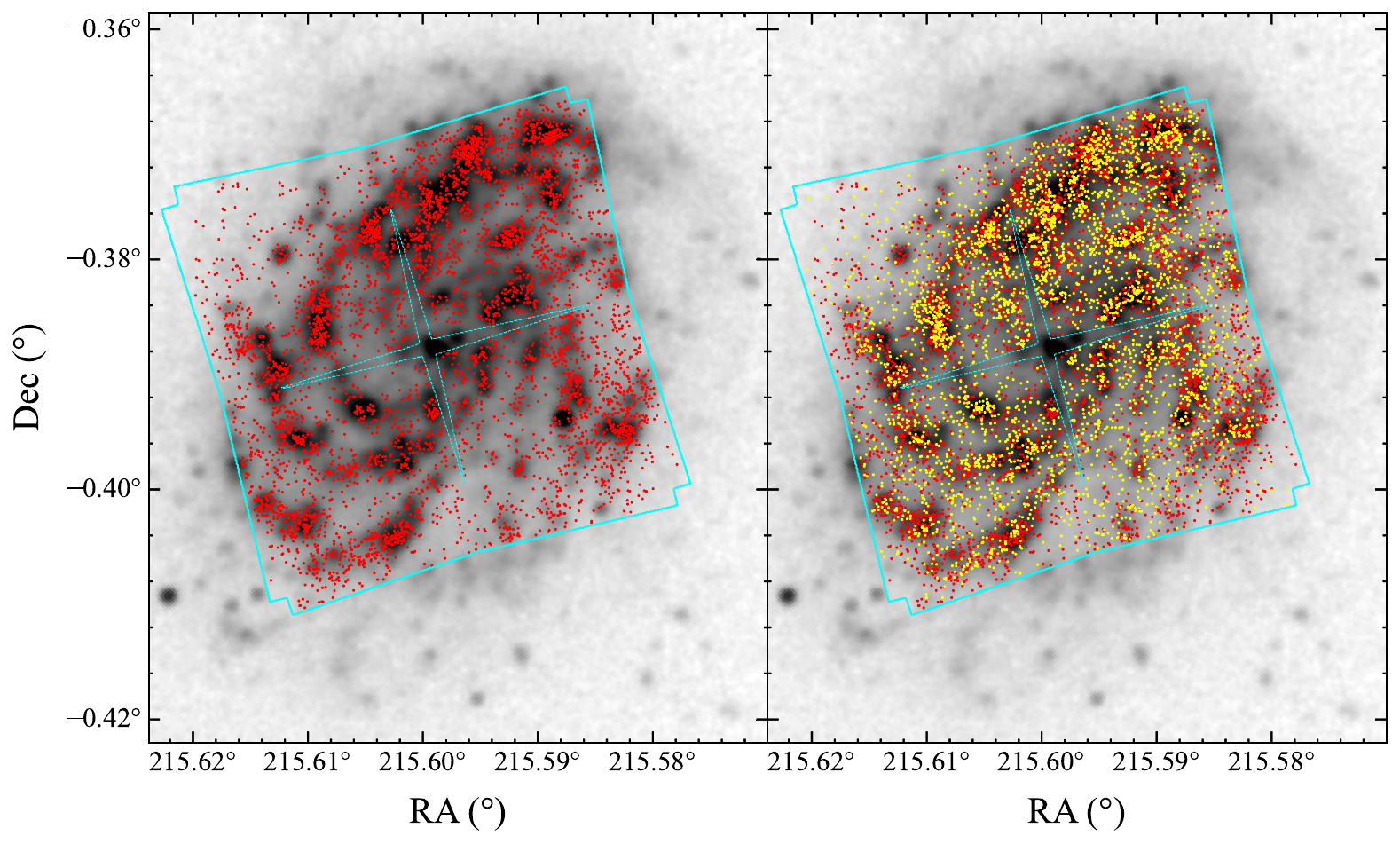}
	\caption{The spatial distribution of RSG candidates, overlaid on a UV image tracing star formation. Left: Our primary sample of RSGs (red dots), selected with a crowding parameter criterion $\le$ 0.5. Right: The primary sample (red) is shown along with an expanded sample (yellow dots), for which the selection criterion is relaxed to crowding $\le$ 1.0. Both the primary and expanded samples clearly trace the galaxy's spiral arms, thereby validating their association with young stellar populations. The cyan lines delineate the observational footprint of JWST/NIRCam. Both samples trace the disk and spiral-arm morphology, while the increase in crowded regions demonstrates that the stringent primary selection is spatially incomplete. The distribution is therefore used to assess completeness and large-scale morphology, not to claim a one-to-one association with current UV emission.}
    \label{figB1}
\end{figure*}

Our primary sample of RSG candidates was selected using a stringent photometric quality cut (crowding $\le 0.5$) to ensure high-precision measurements. However, such strict criteria preferentially remove stars located in the densest, most crowded regions of the galaxy. To quantify this effect, we constructed an expanded sample by relaxing the crowding threshold to $\le 1.0$. As illustrated in Figure \ref{figB1}, this expansion increases the RSG count from 5,310 to 9,250 (a 74.2\% increase). Notably, this more inclusive selection also leads to a significant increase in the number of O-rich AGB stars (a 172.6\% increase), which are intrinsically more numerous in high-density environments.

\subsection{Balmer-decrement check on the UV extinction scale}
\label{sec:balmer_check_discussion}

As described in Section~\ref{sec:muse_data}, archival MUSE spectroscopy covers a subset of the SFC/clump regions in NGC~5584. The UV-$\beta$ method provides the only extinction estimate available for the full SFC/clump sample and is therefore adopted as the fiducial correction for the UV--JWST luminosity comparison. Nevertheless, Balmer decrements provide an independent nebular extinction tracer and have been widely used to estimate dust extinction in star-forming galaxies \citep[e.g.,][]{1992ApJ...388..310K,2006ApJ...642..775M}.
We therefore use the MUSE spectra to check whether the UV-$\beta$ correction has a plausible overall extinction scale for the subset of regions covered by the MUSE fields.

In the Balmer analysis, the stellar continuum is modelled and subtracted using pPXF \citep{2017MNRAS.466..798C} with FSPS stellar population templates \citep{2010ApJ...712..833C} before measuring the H$\alpha$ and H$\beta$ emission-line fluxes. This step is important because underlying stellar absorption near H$\beta$ can bias the Balmer decrement if it is not properly modelled. We assume Case B recombination with an intrinsic H$\alpha$/H$\beta$ ratio of 2.86, appropriate for typical conditions in H\,{\sc ii} regions, with $T_{\rm e}\sim10^4$ K and $n_{\rm e}\sim10^2~{\rm cm^{-3}}$ \citep[e.g.,][]{1989agna.book.....O}. The
nebular colour excess is then computed as
\begin{equation}
E(B-V)_{\rm gas}
=
\frac{2.5}
{k({\rm H}\beta)-k({\rm H}\alpha)}
\log_{10}
\left[
\frac{({\rm H}\alpha/{\rm H}\beta)_{\rm obs}}
{2.86}
\right],
\label{eq:balmer_ebv}
\end{equation}
where $k(\lambda)$ is the extinction curve of \citet{2000ApJ...533..682C}. We then subtract the Milky Way foreground contribution, $E(B-V)_{\rm MW}=A_V/R_V=0.107/3.1=0.0345$ mag, and set any resulting negative internal colour excesses to zero.

Because the Balmer decrement probes nebular extinction, whereas the GALEX FUV and NUV luminosities trace the stellar UV continuum, we convert the internal nebular colour excess to a stellar-continuum colour excess using the Calzetti relation $E(B-V)_{\rm star}=0.44E(B-V)_{\rm gas}$ \citep{2000ApJ...533..682C}. We then evaluate the same extinction curve at the effective wavelengths adopted for the GALEX filters used here,
$\lambda_{\rm FUV}=1548.85$~\AA\ and $\lambda_{\rm NUV}=2303.37$~\AA. This gives $k_{\rm FUV}=10.149$ and $k_{\rm NUV}=8.207$, and hence
\begin{equation}
\begin{aligned}
A_{\rm FUV}
&=
0.44\,k_{\rm FUV}\,E(B-V)_{\rm gas,int}
=
4.465\,E(B-V)_{\rm gas,int},\\
A_{\rm NUV}
&=
0.44\,k_{\rm NUV}\,E(B-V)_{\rm gas,int}
=
3.611\,E(B-V)_{\rm gas,int}.
\end{aligned}
\label{eq:uv_balmer_extinction}
\end{equation}
Equivalently, adopting $k({\rm H}\alpha)=3.326$, these relations correspond to $A_{\rm FUV}=1.343A_{{\rm H}\alpha}$ and $A_{\rm NUV}=1.086A_{{\rm H}\alpha}$ for the GALEX bands used in this work.

For this comparison, we begin with the best available Balmer-decrement measurement for each of the 57 MUSE-covered SFC/clump regions. We exclude only regions for which the Balmer-derived extinction is not physically useful for assessing the UV extinction scale: one region with an exceptionally large Balmer-derived UV extinction (clump 62, $A_{\rm FUV}=5.01$ mag), five regions
with negligible internal Balmer extinction after subtracting the Milky Way foreground contribution ($A_{\rm FUV}\leq0.05$ mag; clumps 2, 4, 5, 59, and 82), and one region without a finite Balmer-derived extinction measurement (clump 87). This leaves a final comparison sample of 50 regions.

For the adopted 50-region comparison sample, the Balmer-derived median extinctions are $A_{\rm FUV}=1.258$ mag and $A_{\rm NUV}= 1.017$ mag, compared with UV-$\beta$ medians of 1.345 and 1.087 mag for the same regions. The corresponding median Balmer-minus-UV-$\beta$ differences are -0.183 and -0.148 mag. This agreement in the median supports the overall scale of the UV-$\beta$ correction, but the weak region-by-region correspondence and substantial scatter show that the methods are not interchangeable. Nebular Balmer emission and the stellar UV continuum are sensitive to different populations, dust geometries, spatial coverages, and timescales; the Balmer comparison is therefore an independent plausibility check rather than a recalibration of individual SFCs.

\begin{figure*}
    \centering
    \includegraphics[width=170mm]{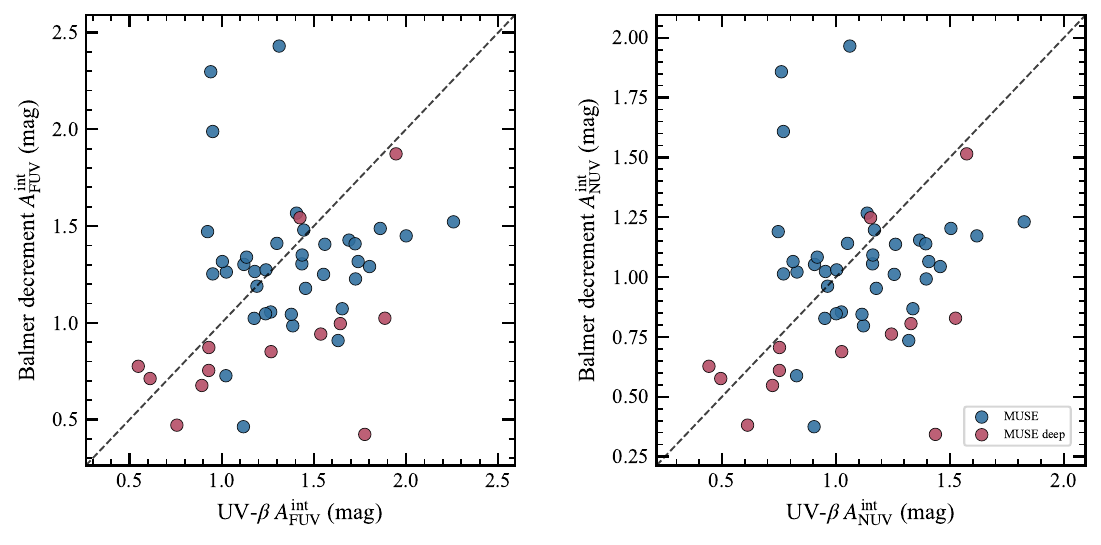}
    \caption{Comparison between the internal UV extinction inferred from the UV-$\beta$ method and that inferred independently from the Balmer decrement for the MUSE-covered SFC/clump regions. The left and right panels show the GALEX FUV and NUV bands, respectively. Blue and red symbols denote regions measured from the MUSE-central and MUSE-deep data cubes. The dashed line marks the one-to-one relation. Before this comparison, we exclude one region with an extreme
    Balmer-derived extinction, five regions with zero or near-zero internal Balmer extinction after foreground subtraction, and one region without a finite Balmer-derived extinction. For the remaining 50 regions, the Balmer and UV-$\beta$ extinction scales agree in the median to within $\sim0.2$ mag, although substantial region-to-region scatter remains.}
    \label{fig:balmer_uvbeta}
\end{figure*}

\subsection{Isochrone fitting methodology and age distribution}
\label{sec:isochron_evolution}

To estimate the characteristic ages of the RSG-rich stellar associations, we performed $\chi^2$ isochrone fitting using MIST models\footnote{\url{https://waps.cfa.harvard.edu/MIST/interp_tracks.html}}. The fitting was carried out for the full set of 106 UV-selected SFC/clump regions, and we then selected a subset of 43 well-populated regions for the final age analysis. These 43 regions each contain at least 12 RSG candidates and yield robust, visually distinguishable fits in the three fitting modes
described below.

In the fiducial MIST fitting, we adopt a metallicity of $[{\rm Fe/H}]=-0.1$ dex. The near-solar gas-phase oxygen abundance of NGC~5584 listed in the HECATE compilation, $12+\log({\rm O/H})=8.75$ \citep{2021MNRAS.506.1896K}, is close to the solar value of $12+\log({\rm O/H})_\odot=8.69$ \citep{2009ARA&A..47..481A} and therefore motivates the exploration of metallicities near the solar value. Because the gas-phase oxygen abundance cannot be translated uniquely into a stellar iron abundance, we tested several near-solar metallicities in the MIST fitting rather than adopting a value directly from the HECATE measurement. Among the metallicities examined, $[{\rm Fe/H}]=-0.1$ dex provided the best overall agreement between the observed CMDs and the model isochrones and was therefore adopted for the fiducial analysis. Fixing the metallicity also reduces the number of free parameters and helps mitigate the partial degeneracy among age, extinction, and metallicity in CMD fitting. The extinction is treated as a free fitting parameter but is constrained to be no lower than the Milky Way foreground extinction, $A_V=0.11$ mag. This lower bound prevents unphysical solutions below the foreground extinction while allowing for additional internal extinction within each UV-selected SFC.

For each region, we fit the F090W versus F090W$-$F150W CMD using both the RSG candidates and the luminous OB candidates. Before the $\chi^2$ calculation, the MIST model grid was filtered to match the stellar populations of interest. RSG models were restricted to $3000<T_{\rm eff}<4500$ K and initial masses $M_{\rm ini}\ge 8\,M_\odot$, while luminous OB candidates were compared with model points satisfying F090W$-$F150W$<0$. The ``global'' fit was obtained by minimizing the combined $\chi^2$ statistic for the RSG and OB candidate populations.

To characterize the luminosity spread of the RSG candidates and the associated age range, we also performed two supplementary RSG-only fits. The ``Top'' fit
uses the brightest 10 per cent of the RSG candidates and traces the younger boundary of the RSG sequence, while the ``Bottom'' fit uses the faintest 10 per cent and traces the older boundary. To maintain statistical robustness, we used at least the five brightest or faintest RSG candidates whenever 10 per cent of the sample contained fewer than five stars. We visually inspected the fitting results for all 106 initial SFC/clump regions and retained the 43 regions in which the global, top, and bottom fits provided a clear and
distinguishable description of the observed CMD morphology. These regions contain $12\le N_{\rm RSG}\le136$ RSG candidates, with a mean of $\langle N_{\rm RSG}\rangle=39$.

The resulting global-fit ages are clustered around
$\log({\rm age/yr})\simeq 7.16$, with a dispersion of $\sim0.06$ dex, and we find no significant large-scale age gradient across the disk. We emphasize that these fits should not be interpreted as proving that every RSG candidate and every OB candidate in a given SFC/clump is strictly coeval. Rather, the fits indicate that the RSG-rich SFC/clump regions are broadly consistent with recent massive-star formation episodes on timescales of several tens of Myr. The UV-bright regions, luminous OB candidates, and RSGs therefore trace related but not identical phases of recent massive-star formation.

\begin{figure*}
    \centering
    \includegraphics[width=85mm]{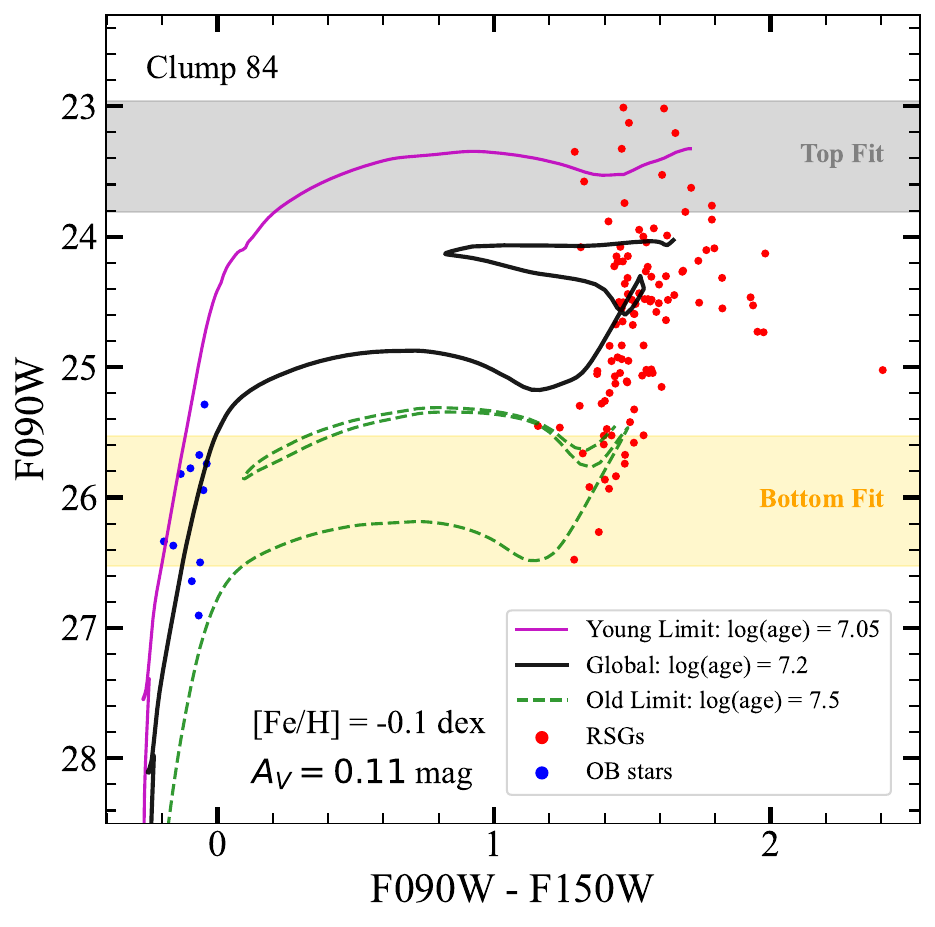}
    \caption{F090W versus F090W$-$F150W CMD for the stellar association in Clump 84. Red and blue points represent the observed RSG candidates and luminous blue candidates, respectively. The stellar populations are fitted with MIST models adopting $[{\rm Fe/H}]= -0.1$ dex, with $A_V$ treated as a fitting parameter subject to a lower limit set by the Milky Way foreground extinction. Three representative fits are shown: the global fit (solid black line), the young boundary constrained by the brightest RSG candidates (solid magenta line), and the old boundary constrained by the faintest RSG candidates (dashed green line). The grey and yellow shaded regions highlight the magnitude ranges used to constrain the young and old age boundaries, respectively.}
    \label{figC1}
\end{figure*}

An example is shown in Figure~\ref{figC1} for Clump 84. The global fit provides a representative age for the association, while the top and bottom fits bracket
the observed luminosity spread of the RSG population. This multi-layered approach is intended to capture the finite age spread and photometric scatter within each SFC/clump region, rather than to assign a unique age to every individual star.

\subsection{C/M-based metallicity as an auxiliary population diagnostic}
\label{sec:mh}

The number ratio of C-AGB to O-AGB stars (the C/M ratio) provides an empirical metallicity diagnostic for stellar populations \citep{2022Univ....8..465R}. Using the empirical relation $[{\rm M/H}]=-0.96\log({\rm C/M})-1.32$ from \citet{2022Univ....8..465R} and our AGB-star counts (Section~\ref{sec:Samples}), we obtain $[{\rm M/H}]=-0.94$ dex for NGC~5584.

This C/M-based estimate is substantially lower than the near-solar metallicity adopted in our fiducial isochrone analysis. Several factors may contribute to this difference. First, the C/M relation was calibrated primarily for relatively metal-poor Local Group systems and may therefore not be directly applicable to the more metal-rich environment of NGC~5584. Second, the two approaches probe different stellar populations: the C/M ratio traces intermediate-age AGB stars, whereas the isochrone analysis considered here is based on young massive stars associated with the UV-selected SFCs. Differences between the metallicities of these populations may arise from the chemical-enrichment history of the galaxy \citep{2002ARA&A..40..487F}. Third, the C/M ratio is sensitive to observational selection effects. Because the O-AGB candidates are generally fainter than the C-AGB candidates in our data, they are more strongly affected by crowding and photometric-quality cuts. Their preferential exclusion would increase the measured C/M ratio and thereby bias the inferred metallicity toward lower values.

The importance of this selection effect is illustrated by the less restrictive crowding-selected sample. Using the AGB counts from this sample, we obtain a higher metallicity of $[{\rm M/H}]=-0.75$ dex because proportionally more O-AGB candidates are recovered. Although this value remains well below the metallicity adopted for the young-star isochrone analysis, the shift demonstrates that the C/M-based estimate depends sensitively on sample selection and differential completeness. Furthermore, the C/M calibration yields an empirical estimate of $[{\rm M/H}]$ for an intermediate-age population, whereas the MIST comparison adopts $[{\rm Fe/H}]=-0.1$ dex for young massive stars; these quantities and populations are not directly equivalent. For these reasons, we do not use the C/M-based metallicity as the fiducial value for the SFC isochrone analysis. Instead, the near-solar abundance of NGC~5584 provides the more relevant prior for the young-star comparison, while the C/M result is retained only as an auxiliary diagnostic of the intermediate-age population and of the sensitivity of empirical metallicity estimates to sample completeness.

\section{Summary}
\label{subsec:summary}

We present a galaxy-wide study of massive-star tracers in the spiral galaxy NGC~5584 by combining high-resolution \textit{JWST}/NIRCam images with \textit{Swift}/UVOT and GALEX ultraviolet data. Our primary sample of evolved massive stars consists of 5,310 RSG candidates selected from the NIRCam CMD. To trace more recent massive-star formation, we define 106 UV-bright SFC/clump regions based on the \textit{Swift}/UVOT UVW2 morphology and measure their GALEX FUV and NUV luminosities within matched apertures.

We find that 82\% of the UV-selected SFC/clump regions coincide with significant RSG overdensities, indicating a strong statistical association between the currently UV-bright SFCs and the evolved massive-star population. This association should not be interpreted as a one-to-one correspondence between individual RSGs and UV emission, because many RSG candidates are located outside the currently UV-bright regions. Instead, these findings suggest that RSG overdensities and UV-selected SFCs trace related but distinct phases of recent massive-star formation across the disk of NGC~5584.

We quantify the connection between the UV-selected SFCs and the near-infrared stellar structures by comparing GALEX FUV/NUV luminosities with \textit{JWST}/NIRCam F090W, F150W, and F277W luminosities measured within matched SFC/clump apertures. After applying the UV-$\beta$ extinction correction, the UV and NIRCam luminosities remain strongly correlated. For the 50 regions covered by MUSE, the median UV-$\beta$ and Balmer-decrement extinction estimates agree to within $\sim0.2$ mag in both FUV and NUV, despite substantial region-to-region scatter. For the 93 clumps with reliable JWST coverage, the Pearson coefficients are $r=0.971$, 0.968, and 0.968 for FUV--F090W, FUV--F150W, and FUV--F277W, respectively, and $r=0.964$, 0.962, and 0.963 for NUV--F090W, NUV--F150W, and NUV--F277W, respectively. These correlations indicate that the UV-selected SFCs and the NIRCam near-infrared stellar light are associated with closely related recent star-forming structures.

Isochrone fitting of the RSG-rich SFC/clump regions with MIST models yields a characteristic age of $\log({\rm age/yr})\approx7.16$ and reveals no significant large-scale age gradient across the disk. We interpret these ages as representative of recent massive-star formation in RSG-rich regions, rather than as evidence that every RSG candidate and every UV-bright or blue massive-star tracer is strictly coeval. Our completeness analysis further indicates that the primary RSG sample is incomplete in the most crowded regions; relaxing the crowding criterion increases the number of RSG candidates by 74\%.

Overall, our results demonstrate the value of combining \textit{JWST} near-infrared images with UV data to connect evolved massive stars, UV-bright SFCs, and near-infrared stellar structures across a spiral galaxy. Our analysis supports a strong statistical connection between RSG overdensities and UV-selected SFCs, while also showing that these tracers represent related but distinct evolutionary phases and spatial scales of recent massive-star formation.

\section*{Acknowledgements}

We are grateful to the anonymous referee for their very helpful comments and suggestions to improve this paper. We are grateful to Drs. Xiaodian Chen, Jun Li, Jiyu Wang, Zhiwen Li and Pinjian Chen for their helpful discussions and suggestions. This work is supported by the NSFC project 12133002, 12373028, and Sichuan Science and Technology Program (Grant No. 2026YFTX0025). S.W. acknowledges the support from the Youth Innovation Promotion Association of the CAS (grant No. 2023065). This work has made use of the data from JWST, $Swift$/UVOT, GALEX and VLT/MUSE.

%%%%%%%%%%%%%%%%%%%%%%%%%%%%%%%%%%%%%%%%%%%%%%%%%%
\section*{Data Availability}
The data underlying this article are available in the article and in its online supplementary material. The data underlying this article will be shared on reasonable request to the corresponding author.
 
%%%%%%%%%%%%%%%%%%%% REFERENCES %%%%%%%%%%%%%%%%%%

% The best way to enter references is to use BibTeX:

\bibliographystyle{mnras}
\bibliography{ref} % if your bibtex file is called example.bib

% Alternatively you could enter them by hand, like this:
% This method is tedious and prone to error if you have lots of references
%\begin{thebibliography}{99}
%\bibitem[\protect\citeauthoryear{Author}{2012}]{Author2012}
%Author A.~N., 2013, Journal of Improbable Astronomy, 1, 1
%\bibitem[\protect\citeauthoryear{Others}{2013}]{Others2013}
%Others S., 2012, Journal of Interesting Stuff, 17, 198
%\end{thebibliography}

% \begin{figure*}
%     \centering
%     \includegraphics[width=170mm]{fig6.pdf}
% 	\caption{F090W versus F090W$-$F150W CMDs for stellar populations in clumps 84 (top), 35 (middle), and 99 (bottom). In each panel, candidate RSGs and OB stars are represented by red and blue dots, respectively. Left panels show the data overlaid with MIST theoretical isochrones of various ages. The right panels show the same data overlaid with MIST stellar evolutionary tracks for various initial masses. All models adopt a solar metallicity ([M/H] = 0). The coloured, dotted lines correspond to the specific ages and masses indicated in the legends.}
%     \label{fig8}
% \end{figure*}
%%%%%%%%%%%%%%%%% APPENDICES %%%%%%%%%%%%%%%%%%%%%

%%%%%%%%%%%%%%%%%%%%%%%%%%%%%%%%%%%%%%%%%%%%%%%%%%

% Don't change these lines
\bsp	% typesetting comment
\label{lastpage}
\end{document}